\documentclass[prl,aps,showpacs,twocolumn]{revtex4}
\usepackage{graphicx}
\usepackage{amsmath}
\usepackage{amsfonts}
\usepackage{amssymb}
\begin{document}

\title{Apparent Persistence Length Renormalization of Bent DNA}
\author{Igor M. Kuli\'{c}}
\altaffiliation{Present address: Department of Physics and
Astronomy, University of Pennsylvania Philadelphia, PA 19104, USA}
\author{Herv\'{e} Mohrbach}
\author{Vladimir Lobaskin}
\author{Rochish Thaokar}
\altaffiliation{Present address: Dept. of Chemical Engineering,
Indian Institute of Technology, Bombay, Mumbai, 400076, India}
\author{Helmut Schiessel}
\altaffiliation{Present address: Instituut-Lorentz, Universiteit
Leiden, Postbus 9506, 2300 RA Leiden, The Netherlands}

\affiliation{Max-Planck-Institute for Polymer Research, Theory
Group, POBox 3148, D 55021 Mainz, Germany}

\date{\today}
\begin{abstract}
We derive the single molecule equation of state (force-extension
relation) for DNA molecules bearing sliding loops and deflection
defects. Analytical results are obtained in the large force limit
by employing an analogy with instantons in quantum mechanical
tunneling problems. The results reveal a remarkable feature of
sliding loops - an apparent strong reduction of the persistence
length. We generalize these results to several other
experimentally interesting situations ranging from rigid
DNA-protein loops to the problem of anchoring deflections in AFM
stretching of semiflexible polymers. Expressions relating the
force-extension measurements to the underlying loop/boundary
deflection geometry are provided and applied to the case of the
GalR-loop complex. The theoretical predictions are complemented
and quantitatively confirmed by MD simulations.
\end{abstract}


\maketitle

In nature DNA is rarely found in its straight ''naked'' state as
usually depicted in the introductory pages of elementary
textbooks. In most ''in vivo'' situations an overwhelming fraction
of DNA is rather strongly configurationally constrained by binding
proteins causing loops, bends and deflections. The advent of
single molecule stretching techniques \cite{Stricketal} has opened
the possibility of measuring the ''equation of state'' of single
tethered DNA molecules in a variety of different conditions
\cite{Stretchingconditions}. While the statistical mechanics of
unconstrained DNA under tension is theoretically well understood
in the framework of the Worm Like Chain model \cite{WLCExp} the
presence of topological constraints like supercoiling
\cite{Strick,BouchMezardEtc} or geometrical constraints like
protein induced kinks and bends
\cite{BruinsmaRudnick,Marko,Kardar} renders analytical results
more difficult. In this letter we expand the repertoire of
analytically solvable ''equations of state'' by deriving the
force-extension relation for a DNA molecule featuring loops and
large deflections in the limit of strong stretching forces (for
the small forces case see \cite{Loop}). The computation is
performed by evaluating quadratic fluctuations around the looped
solution - a non-constant saddle-point of the DNA elastic energy.
The method is essentially analogous to the semiclassical treatment
of tunneling amplitudes in quantum mechanics and instantons in
quantum field theory \cite{PathInt}. After deriving the general
result (that we accompany with MD simulations) we focus on two
interesting experimental applications: the stretching of the GalR
loop complex \cite{Lia}, and of tangentially anchored semiflexible
polymers from a surface in AFM experiments.

\begin{figure}[ptb]
\includegraphics*[width=7cm]{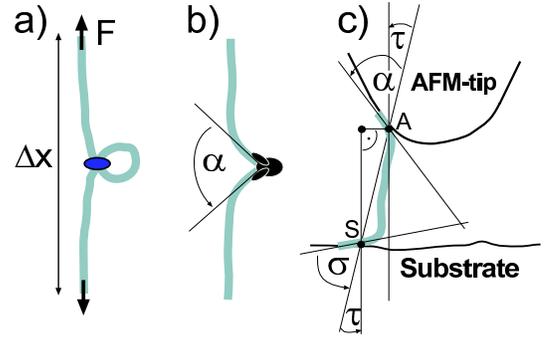}
\caption{Examples for large deflections in a stretched DNA
molecules. a) A freely sliding linker protein stabilizes a DNA
loop. b) A\ rigid ligand with opening angle $\alpha$ causes a kink
in the DNA. c) Tangentially anchored DNA\ stretched from a surface
by an AFM tip. The tilting angle $\tau$ as well as anchoring
angles $\alpha$ and $\sigma$ can strongly affect the elastic
response. }
\label{fig1}
\end{figure}

\textit{Stretching a sliding loop.} In the following we neglect
the DNA twist degree of freedom for it is not constrained from
outside, and no external torsional torques are acting on it. In
this case DNA of length $L$ is described in the continuum limit by
specifying only the unit vector tangent
$\underline{\mathbf{t}}\left(  s\right)  $ to the chain. Here $s$
is the contour length along the DNA with $-L/2<s<L/2$. The chain
is submitted to an external constant force so that the kinetic
plus potential energy of such a chain reads
$E_{0}=\int_{-L/2}^{L/2}\left[  \frac{A}{2}\left(  \frac
{d\underline{\mathbf{t}}}{ds}\right)
^{2}-\mathbf{F}\cdot\underline {\mathbf{t}}\right]  ds$ with the
bending stiffness $A=l_{P}k_{B}T$ where $l_{P}$ is the
orientational persistence length and $k_{B}T$ the thermal energy;
for DNA at room temperature $l_{P}\approx50$ nm \cite{Hagerman}.

In the following we parametrize the tangent as
$\underline{\mathbf{t}}=\left(
\cos\phi\cos\vartheta,\sin\phi\cos\vartheta,\sin\vartheta\right)
$ and put the force along the $x$-axis so that the potential
energy part writes $-F\cos\phi\cos\vartheta$. Note that the angle
$\vartheta$ is measured with respect to the equatorial plane (as
on a globe). This parametrization is necessary to take properly
into account the inextensibility of the chain imposed by the
condition $\underline{\mathbf{t}}^{2}=1$. In the following it is
convenient to introduce the dimensionless contour length
$t=s/\lambda$ with the \textit{deflection length}
$\lambda=\sqrt{A/F}$ \cite{BruinsmaRudnick,OdijkQM}. The latter
becomes the relevant lengthscale characterizing the loss of
orientational correlation in the case of DNA under large tension
(replacing the usual tension-free persistence length
$l_{P}=A/k_{B}T$). In these coordinates the chain energy writes
\begin{equation}
E_{0}=\sqrt{AF}\int_{-L/2\lambda}^{L/2\lambda}\left(  \frac{1}{2}\left(
\dot{\phi}^{2}\cos^{2}\vartheta+\dot{\vartheta}^{2}\right)  -\cos\phi
\cos\vartheta\right)  dt \label{EThetaPHINewCoordinates}%
\end{equation}

The relevant saddle point - the loop in the x--y plane shown in
Fig.~1a - satisfies $\delta E_{0}=0$ with the boundary conditions
$\vartheta _{loop}\left(  \pm L/2\lambda\right)  =0$,
$\phi_{loop}\left(  \pm L/2\lambda\right)  =0$ and $2\pi$. In the
limit of large lengths $L/\lambda\rightarrow\infty$ the loop
solution is the well known ''kink'' solution from theory of
quantum tunneling \cite{PathInt} $\cos\phi
_{loop}=1-2\cosh^{-2}\left(  t\right)  $ and $\vartheta_{loop}=0$
with the loop energy given by $E_{loop}=8\sqrt{FA}-LF$. For large
values of $\sqrt{AF}$ we can expand $E_{0}$ in terms of quadratic
fluctuations $\left( \delta\vartheta,\delta\phi\right)  $ around
the looped saddle point $\left(
\vartheta_{loop},\phi_{loop}\right)  $
\begin{equation}
E_{0}=E_{loop}+\tfrac{\sqrt{AF}}{2}\int\delta\phi\mathbf{\hat{T}}_{\parallel
}\delta\phi dt+\tfrac{\sqrt{AF}}{2}\int\delta\vartheta\mathbf{\hat{T}}_{\perp
}\delta\vartheta dt \label{EThetaPHIWithCotact}%
\end{equation}
with the in- and out-of-plane fluctuation operators $\mathbf{\hat{T}%
}_{\parallel}=\mathbf{-}\partial^{2}/\partial t^{2}+\left(
1-2\cosh ^{-2}t\right)  $ and
$\mathbf{\hat{T}}_{\perp}=\mathbf{-}\partial^{2}/\partial
t^{2}+\left(  1-6\cosh^{-2}t\right)  $ respectively. A closer
inspection of the discrete spectrum of the two operators reveals
the physics behind. $\mathbf{\hat{T}}_{\parallel}$ has a zero
eigenvalue resulting from the translational shift-invariance of
the loop along the chain that costs no energy for
$L/\lambda\rightarrow\infty$. The absence of negative eigenvalues
is in agreement with intuition as in 2D the loop is a
(topologically) stable saddle point. The out-of-plane fluctuation
operator $\mathbf{\hat{T}}_{\perp}$ shows a richer behavior. Again
it possesses a zero mode, this time resulting from rotational
invariance of the problem. More remarkably, in contrast to\
$\mathbf{\hat{T}}_{\parallel}$ the out-of-plane fluctuation
operator $\mathbf{\hat{T}}_{\perp}$ has a negative eigenvalue
$-3$. The latter underlines the intrinsic instability of the loop
in 3D as described by the elastic energy expressions
\ref{EThetaPHINewCoordinates} and \ref{EThetaPHIWithCotact}. This
brings us to the question how to describe the obviously stable
physical situation shown in Fig.~1a. In order to model the action
of a sliding linker \cite{Kardar} we introduce a term accounting
for the DNA self-interaction. In lowest order it can be written as
a function of the perpendicular distance $\Delta z_{c}=\lambda\int_{-t_{c}%
}^{t_{c}}\sin\delta\vartheta\left(  t\right)  dt$ $\approx\lambda\int_{-t_{c}%
}^{t_{c}}\delta\vartheta\left(  t\right)  dt$ of the two overcrossing
DNA\ arms at the contact point $t_{c}\approx1.915$ (resulting from the
crossing condition $t_{c}=2\tanh t_{c}$). The total energy of the chain can
now be written as
\begin{equation}
E_{tot}=E_{0}+V\left(  \lambda\int_{-t_{c}}^{t_{c}}\delta\vartheta\left(
t\right)  dt\right)  \label{Etot}%
\end{equation}
with a short-ranged but otherwise arbitrary attractive interaction potential
$V$ acting at the crossing point. Note that if $V\left(  \Delta z_{c}\right)
$ has a minimum at $\Delta z_{c}=0$ the loop saddle point stays unaffected by
the self-interaction. DNA is known to effectively attract itself in many
solvents despite its strong negative bare charge. Typical situations inducing
DNA self-attraction are poor solvents (like alcohol, small neutral polymers
like PEG) \cite{Bloomfield}, the presence of multivalent counterions (like
CoHex and Spermidine) or small cationic proteins acting as linkers between two
DNA surfaces. Single molecule stretching experiments on DNA condensed with
multivalent counterions performed by several groups \cite{toroids} might bear
loops or related structures like DNA\ toroids \cite{RacquetsRodsToroids}.

The partition function of the system in Fig.~1a can generally be
written as $Q_{loop}=\int_{loop}\delta\left(
\underline{\mathbf{t}}^{2}-1\right)\mathcal{D}^{3}\left[
\underline{\mathbf{t}}\right]  e^{-\beta
E_{tot}\left[\underline{\mathbf{t}}\right]  }$ where $\int_{loop}$
represents the path integral over the functional neighborhood of
the loop solution and the $\delta$-function enforces the chain
inextensibility. This partition function is nothing but the
Euclidean path integral of a quantum particle moving on an unit
sphere under the influence of an external constant force. Using
the $\phi,\vartheta$ parametrisation the partition function can be
rewritten as $Q_{loop}=\int\mathcal{D}\left[ \phi\right]
\mathcal{D}\left[ \vartheta\right]  e^{-\beta E_{tot}\left[
\vartheta,\phi\right] -\ln C_{m}}$ with a metric term
$C_{m}=\exp\left( \int\delta\left( 0\right)  ds\log\left(
\cos\left(  \vartheta\right) \right) \right)  $ resulting from the
inextensibility constraint. It can be shown that this term does
not contribute at the quadratic level of the approximation because
we expand around $\vartheta_{loop}=0$. By virtue of the
decomposition of the in- and out-of-plane fluctuations at the
quadratic level (Eqs.~\ref{EThetaPHIWithCotact} and \ref{Etot})
the partition function can be conveniently factorized
\begin{equation}
Q_{loop}=e^{-\beta E_{loop}}Q_{\parallel}Q_{\perp}^{V} \label{Q}%
\end{equation}
with
\begin{equation}
Q_{\perp}^{V}=\int\frac{d\left(  \Delta z_{c}\right)  }{\lambda}e^{-\beta
V\left(  \Delta z_{c}\right)  }Q_{\perp}\left(  \Delta z_{c}\right)
\end{equation}
and with the in- and out-of-plane plane partition functions
\begin{align}
Q_{\parallel}  &  =\int^{\ast}\mathcal{D}\left[  \delta\phi\right]
e^{-\frac{\beta\sqrt{AF}}{2}\int\delta\phi\mathbf{\hat{T}}_{\parallel}%
\delta\phi dt}\label{QPar}\\
Q_{\perp}  &  =\int^{\ast}\mathcal{D}\left[  \delta\vartheta\right]
\delta\left(  \frac{\Delta z_{c}}{\lambda}-\int_{-t_{c}}^{t_{c}}%
\delta\vartheta dt\right)  e^{-\frac{\beta\sqrt{AF}}{2}\int\delta
\vartheta\mathbf{\hat{T}}_{\perp}\delta\vartheta dt} \label{QPerpen}%
\end{align}
The notation $\int^{\ast}$ reminds that the (translational and rotational)
zero modes of $\mathbf{\hat{T}}_{\parallel}$ and $\mathbf{\hat{T}}_{\perp}$
respectively have to be handled with care. A naive Gaussian integration would
lead to a formal divergence in both cases. After dealing with this problem by
introducing collective coordinates, a method well known from tunneling theory
\cite{PathInt}, and applying the Gelfand-Yaglom method for the computation of
the fluctuation determinant we obtain the in-plane partition function
$Q_{\parallel}=\frac{4}{\pi}\beta LFe^{-\frac{L}{2}\sqrt{\frac{F}{A}}}$.

The computation of $Q_{\perp}$ which follows similar lines of
reasoning requires rewriting the $\delta$-function in
\ref{QPerpen} in its Fourier representation. After introducing
$\Pi_{c}$, the characteristic function of the interval
$[-t_{c},t_{c}]$ (i.e. $\Pi_{c}\left(  t\right)  =1$ for
$t\in\lbrack-t_{c},t_{c}]$ and $=0$ otherwise), we rewrite $\int_{-t_{c}%
}^{t_{c}}\delta\vartheta dt=\int_{-\infty}^{\infty}\Pi_{c}\left(
t\right) \delta\vartheta\left(  t\right)  dt$ as a scalar product.
The path-integral \ref{QPerpen} becomes a Gaussian with a source
term $ik\Pi_{c}$ and can be solved by constructing the Green's
function for $\mathbf{\hat{T}}_{\perp}$. This leads to the result
$Q_{\perp}=8\sqrt{\frac{\Gamma}{\pi}}\left(
\frac{l_{P}}{\lambda}\right)
^{3/2}e^{-\frac{L}{2\lambda}}e^{\Gamma
\frac{l_{P}}{\lambda^{3}}z^{2}}$ with $\Gamma=2/(9t_{c}^{3}-30t_{c}%
)\approx0.35$ a numeric constant. Combining those results we obtain
\begin{equation}
Q_{\perp}^{V}=\frac{8}{\lambda}\sqrt{\frac{\Gamma}{\pi}}\left(  \frac{l_{P}%
}{\lambda}\right)  ^{3/2}e^{-\frac{L}{2\lambda}}\int e^{\Gamma\frac{l_{P}%
}{\lambda^{3}}z^{2}-\beta V\left(  z\right)  }dz\label{QVGeneral}%
\end{equation}
The resulting force extension relation $\left\langle \Delta x\right\rangle
=k_{B}T\frac{\partial}{\partial F}\log Q_{loop}$ writes in leading order%
\begin{align}
\frac{\left\langle \Delta z\right\rangle }{L} &  =1-\frac{1}{2\sqrt{\beta
l_{P}}}\left(  1+8\frac{l_{P}}{L}\right)  \frac{1}{\sqrt{F}}+\frac{9}{4\beta
LF}\nonumber\\
&  +\frac{3\Gamma}{2}\frac{1}{L\lambda}\frac{\int_{-\infty}^{\infty}%
z^{2}e^{\beta\Gamma A^{-1/2}F^{3/2}z^{2}-\beta V\left(  z\right)  }dz}%
{\int_{-\infty}^{\infty}e^{\beta\Gamma A^{-1/2}F^{3/2}z^{2}-\beta V\left(
z\right)  }dz}\label{ForceExtensContact}%
\end{align}
In the limiting case of a very deep potential $V\left(  z\right) $
strongly localized around $z=0$ the last term is negligible and
the force-extension relation becomes independent of the detailed
nature of the contact interaction. If we in addition consider
large forces, the $O\left( 1/\beta LF\right)$-term can be
neglected and the relation
\ref{ForceExtensContact} can be cast in a more illuminating form%
\begin{equation}
\frac{\left\langle \Delta z\right\rangle }{L}\approx1-\frac{1}{2\sqrt{\beta
l_{P}^{app}}\sqrt{F}}\label{Force-Ext-Simple}%
\end{equation}
that resembles the well known loop-free WLC response $\left\langle \Delta
z\right\rangle /L\approx1-\left(  2\sqrt{\beta l_{P}}\sqrt{F}\right)  ^{-1}$
\cite{WLCExp}, but with a strongly renormalized apparent ''persistence
length'' $l_{P}^{app}$ :
\begin{equation}
l_{P}^{app}=l_{p}\left(  1+8\frac{l_{p}}{L}\right)  ^{-2}\label{LPapp}%
\end{equation}
Equations \ref{Force-Ext-Simple} and \ref{LPapp} show that one has
to be very cautious when interpreting experimental stretching data
in terms of persistence length and stiffness. The presence of a
loop modifies the elastic response of the chain in such a manner
that the persistence length can appear effectively reduced as
stated in \ref{LPapp}. For a single loop this is obviously a
finite size effect involving the scaled total length $L/l_{p}.$
But the effect remains significant over a large range of
parameters: for $l_{P}/L=0.1$ one finds
$l_{P}^{app}\approx0.31l_{P}$ whereas for
$l_{P}/L=0.02$ there is still a remarkable effect, namely $l_{P}^{app}%
\approx0.74l_{P}$.

\begin{figure}[ptb]
\includegraphics*[width=7cm]{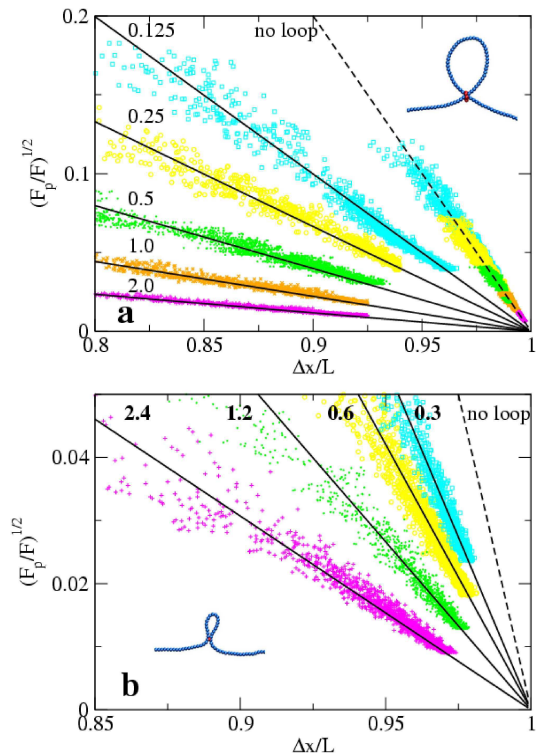} \caption{ a) Stretching DNA looped
with a sliding linker: Comparison of MD simulation results
(datapoints) with theoretical prediction (solid lines) for the
force-extension relation as given by Eq.~\ref{Force-Ext-Simple}.
The force is measured in units of $F_{P}=k_{B}T/l_{P}$. The
different slopes correspond to different values of $l_{P}/L$. The
dashed line shows free chain (no loop) behavior. b) Stretching of
a GalR loop complex: MD simulation (dots) vs. theory, Eq.~\ref{LappGalR}
(solid lines) for various values of $l_{P}/L$.}%
\label{fig2}
\end{figure}

To check the prediction of Eqs.~\ref{Force-Ext-Simple} and
\ref{LPapp} we performed molecular dynamics (MD) simulations
\cite{footnote1}. As seen from Fig.~2a the theoretical predictions
quantitatively agree with the simulation results.

\textit{Applications.} In order to generalize the asymptotic
(large $F$) relations \ref{Force-Ext-Simple} and \ref{LPapp} to
other interesting situations we can rewrite them in a more
illuminating fashion in terms of the
three relevant lengths $L,l_{P}$ and $\lambda$%
\begin{equation}
\frac{\left\langle \Delta z\right\rangle }{L}\approx1-\left(  \frac{\lambda
}{2l_{P}}+\frac{4\lambda}{L}\right)  \label{ForcExtNice}%
\end{equation}
The first term in the brackets is the length-loss by thermal
fluctuations. Because of the small size of the loop in the large
$F$ limit ($\lambda\ll L$) it comes as no surprise that the term
$\lambda/2l_{P}$ coincides with the thermal fluctuation
contribution of a \textit{loop-free chain}. On the other hand the
second term in the bracket $4\lambda/L$ describes the loss of
length due to the (loop induced) elastic deflection. This
asymptotic decomposition of the two contributions in
\ref{ForcExtNice} leads to immediate generalizations. After short
inspection it is easy to see that any strongly localized
\cite{footnote2} DNA deflection (like in Fig.1b and 1c) can be
appropriately continued and mapped piecewise onto fractions of the
full loop solution, Fig.1a. By this reasoning a deflection angle
of any size (between $0$ and $2\pi$) can be immediately related to
the length loss in complete analogy to the loop case.

As a first example consider the stretching of rigid DNA-protein
complexes which come as large kinks or fixed angle loops, cf.
Fig.1b. A prominent example is the GalR repressosor complex that
was studied in single molecule experiments \cite{Lia}. Following
the upper reasoning one obtains the same force-extension relation
as in \ref{Force-Ext-Simple} but with an apparent persistence
length given as a function of the opening angle $\alpha$ of the
complex:
\begin{equation}
l_{P}^{app}=\frac{l_{p}}{\left(  1+8\frac{l_{p}}{L}\left(  1-\cos\left(
\frac{\pi-\alpha}{4}\right)  \right)  \right)  ^{2}} \label{LappGalR}%
\end{equation}
For the GalR complex the opening angle is not known but there are
some indications for the ''anti-parallel'' configuration, i.e.,
$\alpha\approx0$ \cite{Lia,Geanacopoulos}. With
\ref{Force-Ext-Simple} and \ref{LappGalR} at hand one can predict
the loss of length due to the presence of a GalR complex with the
conjectured angle $\alpha=0.$ For the force $F=0.88$ pN applied in
the experiment \cite{Lia} with a loopsize of 38 nm \cite{Lia},
Eqs.~\ref{Force-Ext-Simple} and \ref{LappGalR} predict a loss of
length of 56 nm (with the DNA hidden inside the loop included).
Remarkably, the experimental value by Lia et al. \cite{Lia} is 55
nm $\pm 5$ nm. The latter result together with
\ref{Force-Ext-Simple} and \ref{LappGalR} gives convincing
evidence for the ''anti-parallel'' loop model. An independent
check of \ref{Force-Ext-Simple} and \ref{LappGalR} is provided by
MD simulations (shown in Fig.2b) which were performed for various
ratios of $l_{p}/L$. In conclusion, Eq.~\ref{Force-Ext-Simple}
enables one to directly measure the angle $\alpha$ - an important
feature of the DNA-protein complex geometry.

A second application concerns AFM\ stretching experiments of semiflexible
polymers \cite{Janshoff}. A stable anchoring can be achieved when the polymer
is tangentially attached at its two ends as in Fig 1c. Force extension data in
such a setup have to be interpreted with care. The boundary anchoring angles
at the AFM tip as well as at the surface - $\alpha$ and $\sigma$ respectively
- can significantly alter the measured apparent persistence length, a fact
which has been overlooked before. A more trivial effect of a tilting angle
$\tau$\ in Fig. 1c (between the line of contact-points and the force
direction) can alter the result in addition \cite{Rief}. A simple calculation
for large forces $F$ again gives the same functional relation as in
\ref{Force-Ext-Simple} with an ''apparent persistence length'' given by%
\begin{equation}
l_{P}^{app}=\frac{l_{p}\cos\left(  \tau\right)  }{\left(  1+8\frac{l_{p}}%
{L}\left(  1-\frac{1}{2}\left(  \cos\frac{\alpha}{2}+\cos\frac{\sigma}%
{2}\right)  \right)  \right)  ^{2}} \label{LAppAFM}%
\end{equation}
While the angle $\tau$ can be completely canceled by shifting the
tip in the lateral direction, the influence of $\alpha$ and
$\sigma$ cannot be fully eliminated by this procedure. This means
that for short to intermediate sized semiflexible polymers
($l_{P}/L\sim1-1/30$) the influence of boundary conditions has to
be taken into account through Eq.~\ref{LAppAFM}.

In conclusion, the force-extension relation of looped DNA which we derived in
the limit of strong stretching force can be generalized to DNA featuring large
deflections. In particular our approach provides an analytical expression for
the force-extension response of DNA-bending proteins which can give important
information with regard to the DNA-protein complex geometry. Large deflections
due to the anchoring at the AFM tip and at the substrate also affect the
measured persistence length for not too long molecules. Finally these effects
might be related to the strong reduction of the persistence length found in
condensed DNA stretching experiments by Baumann et. al. \cite{toroids}.

The authors thank Nikhil Gunari, Andreas Janshoff and Phil Nelson for useful discussions.


\begin{thebibliography}{99}
\bibitem{footnote1} Langevin MD simulation of a
semi-flexible bead-spring chain with up to 200 beads; the
stretching force applied at the end beads; the contact potential
was realized by a flexible sliding linker i.e. closed bead-spring
chain \cite{Kardar}.
\bibitem{footnote2} By strongly localized we mean a deflection that
decays to zero on a lengthscale much shorter than the whole
molecule.
\end{thebibliography}
\end{document}